\documentclass{jfm}

\usepackage{natbib}
\usepackage{tikz}
\usepackage{gensymb}

\usepackage{color}

\usepackage{breakcites}
\usepackage{microtype}

\ifCUPmtlplainloaded \else
  \checkfont{eurm10}
  \iffontfound
    \IfFileExists{upmath.sty}
      {\typeout{^^JFound AMS Euler Roman fonts on the system,
                   using the 'upmath' package.^^J}%
       \usepackage{upmath}}
      {\typeout{^^JFound AMS Euler Roman fonts on the system, but you
                   dont seem to have the}%
       \typeout{'upmath' package installed. JFM.cls can take advantage
                 of these fonts,^^Jif you use 'upmath' package.^^J}%
      }
  \else
  \fi
\fi

\ifCUPmtlplainloaded \else
  \checkfont{msam10}
  \iffontfound
    \IfFileExists{amssymb.sty}
      {\typeout{^^JFound AMS Symbol fonts on the system, using the
                'amssymb' package.^^J}%
       \usepackage{amssymb}%
         
         \let\geq=\geqslant
      }{}
  \fi
\fi

\ifCUPmtlplainloaded \else
  \IfFileExists{amsbsy.sty}
    {\typeout{^^JFound the 'amsbsy' package on the system, using it.^^J}%
     \usepackage{amsbsy}}
    {}
\fi

\newcommand\Rey{\mbox{\textit{Re}}}

\newsavebox{\astrutbox}
\sbox{\astrutbox}{\rule[-5pt]{0pt}{20pt}}

\title[Localised periodic orbits in plane Poiseuille flow]{Streamwise and doubly-localised periodic orbits in plane Poiseuille flow}

\author[S. Zammert and B. Eckhardt]%
{S\ls T\ls E\ls F\ls A\ls N \ns Z\ls A\ls M\ls M\ls E\ls R\ls T $^{1}$ \thanks{Email address for correspondence: stefan.zammert@physik.uni-marburg.de}
\and B\ls R\ls U\ls N\ls O\ns E\ls C\ls K\ls H\ls A\ls R\ls D\ls T$^{1,2}$}

\affiliation{$^{1}$ Fachbereich Physik, Philipps-Universit\"at Marburg, Renthof 6, D-35032 Marburg, Germany\\[\affilskip]
$^{2}$ J.M. Burgerscentrum, Delft University of Technology, Mekelweg 2, 2628 CD Delft, The Netherlands}

\begin{document}

\maketitle

\begin{abstract}
We study localised exact coherent structures in plane Poiseuille flow that are relative periodic 
orbits.  They are obtained from extended states in smaller periodically continued domains,
by increasing the length to obtain streamwise localisation and then by increasing the width to achieve 
spanwise localisation. The states maintain the travelling wave structure of the extended states, which
is then modulated by a localised envelope on larger scales. In the streamwise direction, the envelope
shows exponential localisation, with different exponents on the upstream and downstream sides.
The upstream exponent increases linearly with Reynolds number $\Rey$, but the downstream exponent is
essentially independent of $\Rey$. In the spanwise direction the decay is compatible with
a power-law localisation.  As the width 
increases the localised state undergoes further bifurcations which add additional unstable directions,
so that the edge state, the relative attractor on the boundary between the laminar and turbulent motions, in the system becomes chaotic.
\end{abstract}

%
\section{Introduction}
\begin{sloppypar}
The study of coherent structures in small periodic domains, often referred to as 'minimal flow units', has provided 
considerable insight into the phase-space structure and the transition dynamics of shear flows without linear
instabilities of the laminar state, such as pipe flow or plane Couette flow and various boundary layers (e.g. \citet{Kreilos2012} and references
therein). For an understanding of the fascinating spatio-temporal dynamics in the transition region, where intriguing
patterns of alternating laminar and turbulent dynamics \citep{Barkley2005,Duguet2010b} 
and a complicated evolution that has been linked to directed 
percolation \citep{Manneville2009,Moxey2010,Avila2011} can be observed, it is necessary to investigate spatially extended domains and localised states. 
The interest in localised solutions also arises from the possibility of 
using them as building blocks for more complicated spatial patterns, such as the turbulent spots observed in 
plane Couette or plane Poiseuille flow in various experimental 
\citep[e.g.][]{Carlson1982,Dauchot1995,Hegseth1996,Lemoult2013} and numerical 
\citep[e.g.][]{Henningson1987,Lundbladh1991,Schumacher2001} studies.
\end{sloppypar}

The spatially extended exact coherent structures arise in the form of 
stationary states without any time-dependence, travelling waves where a pattern moves downstream with a fixed 
speed, or relative periodic orbits that return to the initial pattern except for a displacement
\citep{Nagata1990,Nagata1997,Ehrenstein1991,Schmiegel1999,Wang2007,Gibson2009}. 
Spanwise localised exact states have been 
identified by \citet{Schneider2010a,Schneider2010}, \citet{Gibson2014} 
and \citet{Zammert2013}  for plane Couette and plane Poiseuille flow.  
Streamwise localised exact solutions have been described for the case of 2D plane Poiseuille flow 
 \citep{Price1993} and for pipe flow \citep{Avila2013,Chantry2013}. 
Spanwise localised states in plane Poiseuille flow show a complicated temporal dynamics \citep{Zammert2013} 
that has also been documented  for  the asymptotic suction boundary layer  by  \citet{Khapko2013,Khapko2013a}. A state in plane Couette flow that is localised in both 
the spanwise and streamwise directions has been identified recently by \citet{Brand2014}. 

In this paper we present coherent structures for plane Poiseuille flow that are localised in the streamwise direction
and in streamwise and spanwise direction. We find them using the method of edge tracking  \citep{Skufca2006}
in small periodic  domains, and then continue them first to longer and then 
also to wider domains. We begin with a discussion of the extended states in narrow domains in section 2, 
followed by studies of streamwise localised states in section 3 and spanwise and streamwise localised
structures in section 4. Conclusions are given in section 5.

\section{The edge state in short domains -  a traveling wave}

We study the incompressible plane Poiseuille flow (PPF), the pressure driven flow between two infinitely extended 
parallel plates. 
With the $x$-axis along the flow direction,
the plates are parallel to the $x$-$z$ plane at $y=\pm h$. 
The Reynolds number is based on $h$, the laminar  centreline velocity $U_{0}$ and the kinematic viscosity $\nu$, so that
$Re=U_{0}h/\nu$ and the laminar non-dimensional profile becomes $U(y)=(1-y^{2})$. 
In all our simulations constant mass flux is imposed. The velocity fields used in the following are
the deviations from the laminar profile, denoted  $\textbf{u}=(u,v,w)$, where $u$, $v$ and  $w$ are the 
streamwise, wall-normal and spanwise velocity components, respectively.

The numerical simulations are based on the spectral code \textit{channelflow}, developed and maintained 
by \citet{J.F.Gibson2012}.
The package provides a Newton method \citep{Viswanath2007} for searching for exact solutions 
as well as tools for continuation and stability analysis. 
We adapted the channelflow-code to work with parallel FFTW (OpenMP) 
and replaced the \textit{Octave}-library used for the linear algebra routines in the Newton method 
and the eigenvalue calculations by the \textit{Eigen}-package \citep*{eigenweb}.
The method of the edge tracking algorithm is described, e.g. by 
\citet{Toh2003}, \citet{Skufca2006}, \citet{Schneider2008} and \citet{Dijkstra2013}.

We started off with edge tracking from a random initial condition in a small periodic domain with 
streamwise length $L_{x}$ of $2\pi$ and spanwise width $L_{z}$ of $2\pi$ with a numerical resolution of 
$N_{x}\times N_{y} \times N_{z}=32\times65\times48$ and a Reynolds number  of 1400. 
We checked our resolution  by comparing to a higher one of $N_{x}\times N_{y} \times N_{z}=80\times97\times112$.
For this \textit{Re} plane Poiseuille flow shows persistent turbulence although it is far below the critical Reynolds number of 5772 \citep{Orszag1971}
that follows from linear stability theory. Note that on account of the domain length of $2\pi$ the actual critical Reynolds number in this domain is 5815. 
Edge tracking usually converges quickly to one travelling wave, referred to as
$TW_{E}$ in the following. This state 
has two symmetries: a mirror symmetry with respect to the mid-plane, and a shift-and-reflect symmetry in the
spanwise direction,
\begin{eqnarray}
s_{y}: [u,v,w](x,y,z)=[u,-v,w](x,-y,z)\,,\\
s_{z}\tau_{x}: [u,v,w](x,y,z)=[u,v,-w](x+ L_{x}/2,y,-z)\,.
\end{eqnarray}
The state is dominated by a strong low-speed streak in the mid-plane and pairs of vortices at the top and 
bottom plate. 
The travelling waves has the same symmetry as TW1-1 from \citet{Gibson2014} and W01 from \citet{Waleffe2001} and \citet{Nagata2013a}.

Stability analysis of the travelling wave in the full space without any symmetry restriction 
shows that it has one unstable eigenvalue for $510<Re<5850$ so that its stable manifold is of co-dimension one. 
A stability analysis of this periodic state in longer domains shows that for $L_{x}=4\pi$ the state already 
has an additional pair of unstable complex-conjugated eigenvalues for $Re<1785$. Further doubling of the domain size adds 
more unstable directions, so that, e.g., for $L_{x}=8\pi$ and $Re=1400$ the wave has 5 unstable eigenvalues. 
The long-wavelength instabilities in the larger domains are precursors to 
localisation in the  spanwise \citep{Melnikov2014} and streamwise directions \citep{Chantry2013}.

\section{Streamwise localised periodic orbits in long domains}
\begin{figure*}
\centering
\includegraphics[width=0.95\textwidth]{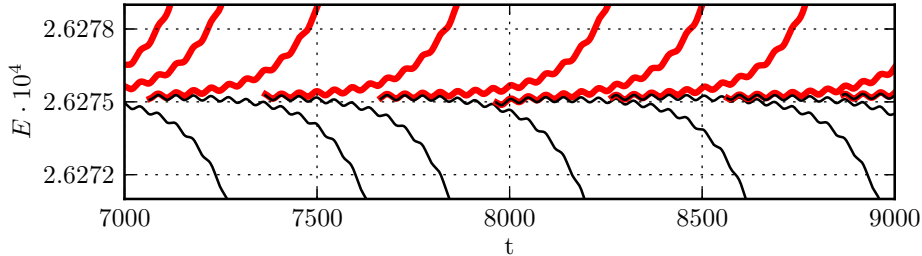}
\caption{Edge tracking in a computational domain of length $L_{x}=32\pi$ and width $L_{z}=2\pi$ 
for $\Rey= 1400$.
Shown are the energy densities of trajectories that turn turbulent (red thick) and laminar (black thin), respectively.
The edge state bracketed by these trajectories oscillates periodically in energy.
}
\label{fig_ETR1400}       
\end{figure*}

In a longer domain of length $32 \pi$ but  with the same width of $2\pi$ and at Reynolds 
number $\Rey=1400$, edge tracking converges to a state that at first glance looks like a state of constant energy density,
\begin{eqnarray}
E(\textbf{u}) = \frac{1}{4L_{x}L_{z}}\int \textbf{u}^{2} dx dy dz. 
\end{eqnarray}
However, closer inspection of the time trace in figure \ref{fig_ETR1400}
reveals that it is not constant but shows a regular oscillation with an amplitude of order $10^{-8}$. 
This oscillation is not a numerical artefact but reflects properties of the edge state, as we now show.

\begin{figure*}
\centering
\includegraphics[]{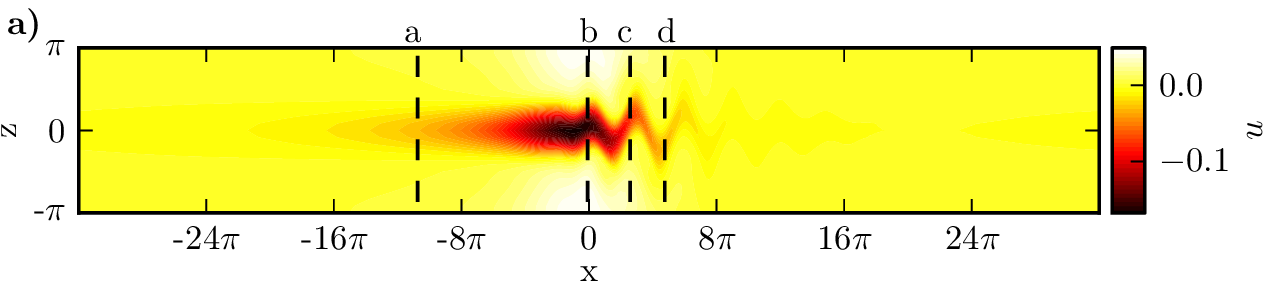}
\\
\includegraphics[]{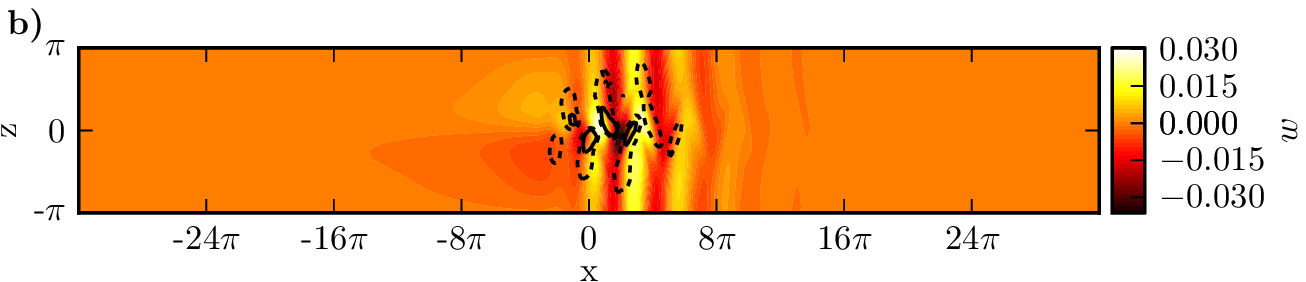}
\caption{Instantaneous velocities in the mid-plane for the edge state $PO_{E}$ at $\Rey=1400$ at the time of minimal energy.
Shown are the (a) streamwise and (b) spanwise velocities.
The black lines in (a) mark the positions of the spanwise wall-normal cross sections 
in \ref{fig_YZplanePOEandASY} (a)-(d).
The solid and dashed lines in (b) are iso-contours of the Q-vortex criterion \citep{Hussain95} at levels of 0.001 
and 0.0001, respectively. 
The direction of the flow is from left to right.
\label{fig_PO_XZplanesR1410}
\label{fig_POE_XZplane}}
\end{figure*}

\begin{figure}
\includegraphics[]{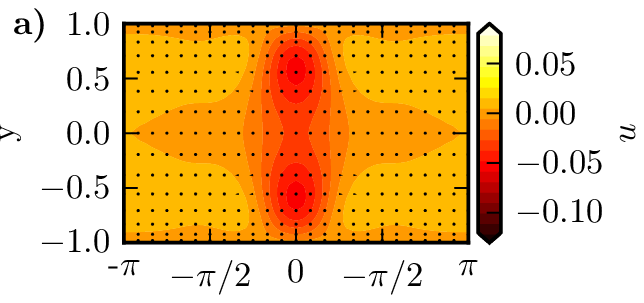}
\includegraphics[]{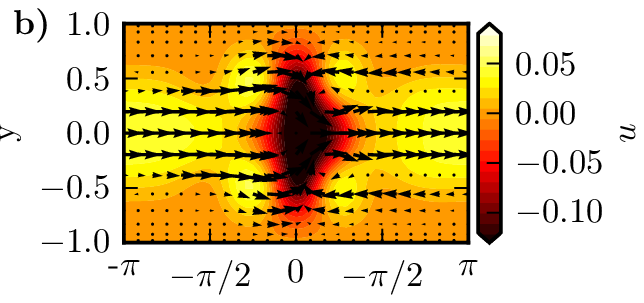}
\\
\includegraphics[]{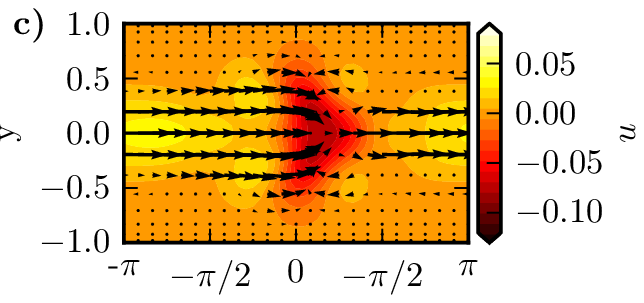}
\includegraphics[]{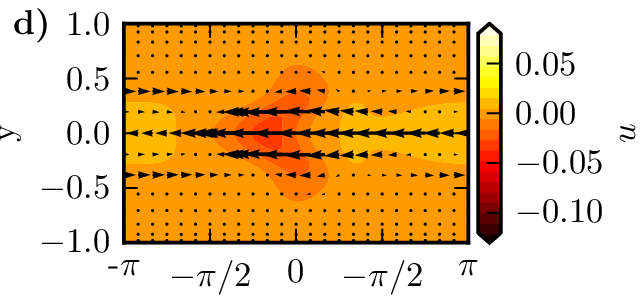}
\\
\includegraphics[]{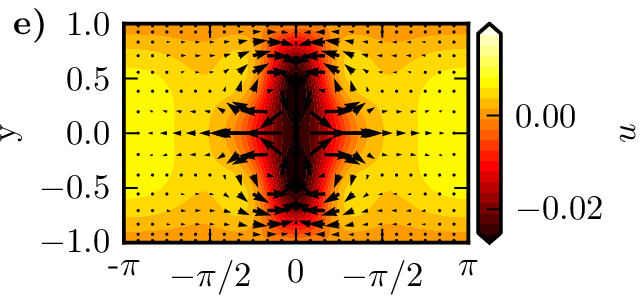}
\includegraphics[]{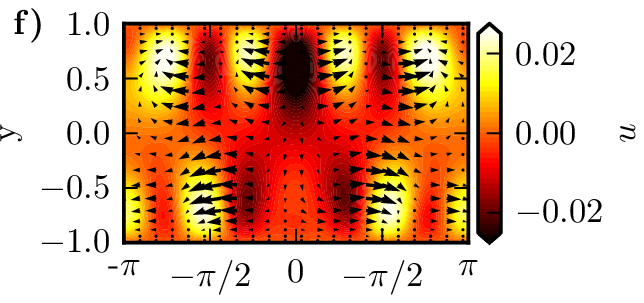}
\caption{(a) - (d) Instantaneous cross sections of the edge state $PO_{E}$ at $Re=1400$ 
at the streamwise positions indicated 
in figure \ref{fig_POE_XZplane}(b).
The in-plane components of the velocity are indicated by arrows and the streamwise component is colour coded. 
(e) and (f) show the streamwise-averages 
of  $PO_{E}$ at $Re=1400$ and of the orbit that bifurcates from it, $PO_{asy}$, at $Re=1625$, respectively. \label{fig_YZplanePOEandASY}  }    
\end{figure}

\begin{figure}
\centering
\includegraphics[]{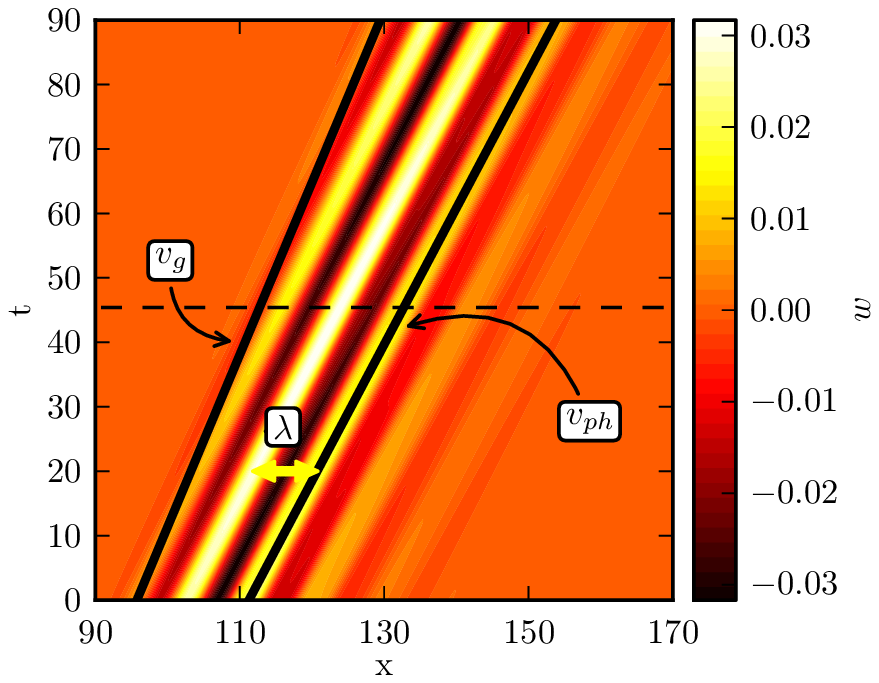}
\caption{Space-time-display of the spanwise velocity of the periodic orbit $PO_{E}$ in the mid-plane $y=0$ at
at $z=0$ for $\Rey=1400$. The solid black lines indicate the group velocity $v_{g}$ and 
the phase velocity $v_{ph}$ for one of the maxima. The wavelength  $\lambda$ of the internal
modulations varies along the state. The dashed black line marks one period $T$ in energy.
\label{fig_ETR1400_spatioTemp}
}
\end{figure}

Although the initial velocity field is spatially non-localised, the flow state obtained by edge tracking is 
localised in the streamwise direction. 
Using this state as an initial condition in a Newton method, ther is a quick convergence to a streamwise localised 
relative periodic orbit, referred to as $PO_{E}$ in the following. 
For further study we transfer the state to a computational domain of length $64 \pi$,  which
is possible because of its streamwise localisation.
We use a resolution of $768\times65\times48$ and check our results with a higher spanwise resolution of $N_{z}=80$. 
The streamwise and the spanwise velocity components in the mid-plane are shown in figure \ref{fig_POE_XZplane}.
Cross-sections for Re$=1400$ for different streamwise positions are shown in 
figure \ref{fig_YZplanePOEandASY} (a) - (d) and the streamwise averaged velocity is shown in (e).
The images reveal that the orbit has a mirror symmetry ($s_{y}$) but no 
$s_{z}\tau_{x}$-symmetry and is dominated by a strong narrow low-speed streak and 
a weak and extended high-speed streak. It is therefore very similar to the travelling wave $TW_{E}$. 
In particular, the streamwise-averaged flow for $TW_{E}$ is close 
to the one shown in figure \ref{fig_YZplanePOEandASY} (e). 

The complex spatial propagation pattern of the wave can be seen in figure \ref{fig_ETR1400_spatioTemp} where the spanwise velocity in the mid-plane at $z=0$  is plotted versus time for $\Rey=1400$.
The total energy of the state is periodic with a period $T=45.402$, but the state needs twice this time to return
in shape, apart from a downstream shift. 
After a time $T$  it returns apart from a symmetry operation $s_{z}: [u,v,w](x,y,z)=[u,v,-w](x,y,-z)$ and a downstream
shift that is half the one after $2T$. The figure clearly shows a group velocity $v_g$ for the envelope of the state, 
and a phase velocity $v_{ph}$ for the underlying structures. 
The group velocity $v_{g}$ can be calculated by dividing the distance travelled over two periods by $2T$. 
For $\Rey=1400$ one obtains $v_{g}=0.8753$, indicated by the line at the upstream end of the state in 
figure \ref{fig_ETR1400_spatioTemp}.
The structures underneath the envelope move with a different velocity $v_{ph}$, which can be read off from the slope of the maxima. 
However, the wavelength $\lambda$ of the spanwise modulations varies 
slightly with position and is slightly larger at the front
than in the centre and towards the end of the state. Therefore, the velocity of each maximum also varies slightly, so
that the phase velocity depends on the position within the state. 

Using a continuation method \citep[see e.g.][]{Dijkstra2013} it is easy to track the state in $\Rey$. It turns out that the periodic
orbit exists down to $\Rey$ $\approx 1038$, where it is created in a saddle-node bifurcation. 
Furthermore, it is also possible to identify the upper branch of the periodic orbit. 
This upper branch is also localised and has multiple unstable directions.

\begin{figure}
\centering
\includegraphics[]{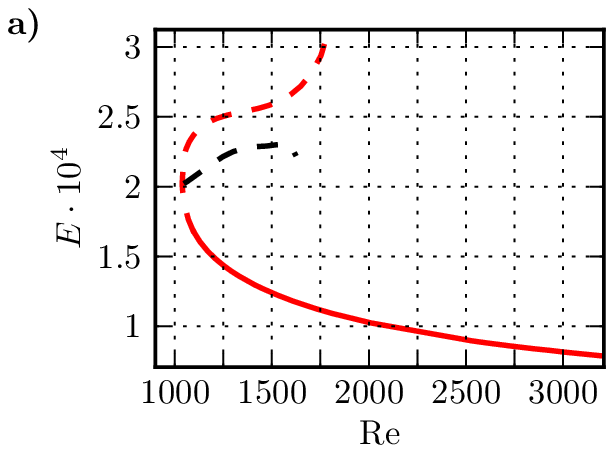}
\includegraphics[]{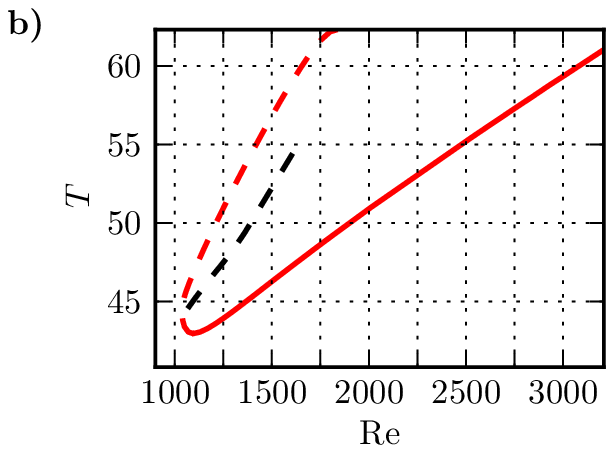}
\caption{Bifurcation diagram for the coherent structures. (a) Minimum energy of the localised periodic orbit $PO_{E}$ (red) 
and the asymmetric orbit $PO_{asy}$ (black) vs  $Re$. In both cases the minimum in energy over the period $T$ is shown.
The stability of the orbit is indicated by the linetype, solid for a single unstable direction, and dashed for more than one.
(b)  Variation of the periods in energy with $\Rey$. 
\label{fig_BifDiag}}  
\end{figure}

A stability analysis of the lower branch state shows that for $\Rey>1100$ it has one unstable direction.
Therefore, for these Reynolds numbers the state is an edge state whose stable manifold 
can separate the state space into two parts. 
For lower values of $\Rey$ the periodic orbit has more than one unstable direction. 
The bifurcation near $\Rey\approx1100$ is a Sacker-Neimark bifurcation 
\citep{Kuznetsov1998} that breaks the $s_{y}$ symmetry. 
This bifurcation is followed by further bifurcations resulting in eight unstable directions for the lower branch. 
One of the bifurcations of the lower branch is a  pitchfork bifurcation that breaks the $s_{y}$ symmetry and 
creates an asymmetric periodic orbit ($PO_{asy}$).
The image of the state in figure \ref{fig_YZplanePOEandASY} (f) shows that its internal dynamics is 
more complex than that of $PO_{E}$. 
The bifurcations and some properties of the states are summarised in figure \ref{fig_BifDiag}.

\begin{figure}
\includegraphics[]{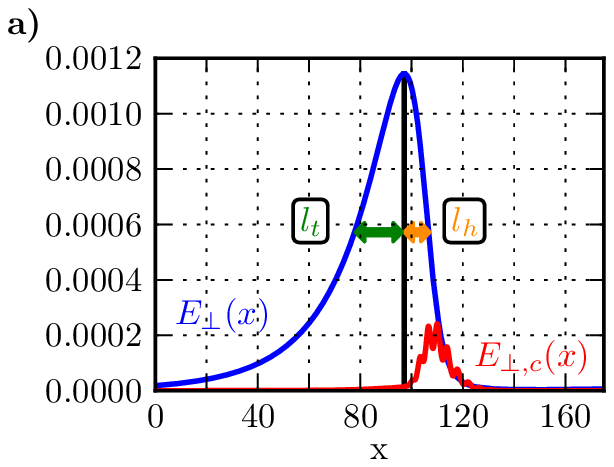} 
\includegraphics[]{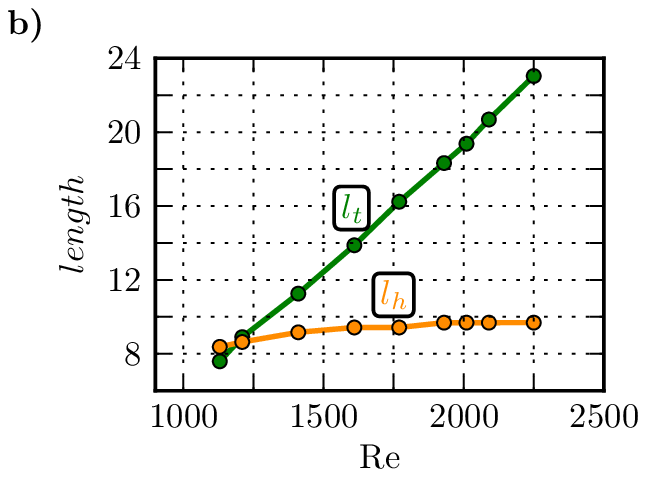}
\caption{Energy profiles of the localised state. (a) Densities of total (red) and the cross flow (blue) energy 
for the periodic orbit at $Re=2010$ along the flow direction, which is from left to right. 
The cross flow energy density is multiplied by a factor of 10.
(b) Dowstream  $l_{h}$ (orange) and upstream $l_{t}$ (green) distances from the maximum to half the maximal
values in energy. 
}
\label{fig_DensityAndSize}       
\end{figure}

Information about the localisation properties can be extracted from the streamwise variation of the
energy density of the deviation from the laminar flow,
\begin{equation}
 E_{\perp}(x)=\frac{1}{4 L_{z}}\int_{0}^{L_{z}}\int_{-1}^{1} \textbf{u}^{2} \ dy dz ,
\end{equation}
and the density of the cross-flow energy 
\begin{equation}
 E_{\perp,c}(x)=\frac{1}{4 L_{z}}\int_{0}^{L_{z}}\int_{-1}^{1} v^{2}+w^{2} \ dy dz .
 \label{eq_DensityCross}
\end{equation}
The energy densities for the periodic orbit at $Re=2010$ (at times of minimal energy) are shown in figure 
\ref{fig_DensityAndSize}(a). One can identify a small region with relatively high cross-flow energy at the front of the state.
In this region the cross-flow draws energy from the laminar profile and transfers it into streamwise velocity, 
which then drives streaks and causes a steep increase of the total energy density at the front of the state. 
The energy in the streamwise components has its maximum at a position in the tail where the 
cross-flow energy is already very low again.  In the absence of cross-flow motion the  streaks are dampened
by viscosity only, which results in the long tail of the state.

Based on the energy density $E_{\perp}(x)$ one can introduce two characteristic length scales for $PO_{E}$,
associated with the extension in the downstream and upstream direction. Starting from the maximum in energy,
one can determine the distances to the locations where the energy density has dropped to half its maximum.
They are denoted $l_{t}$ and $l_{h}$ for the upstream (tail) and downstream (head) sides, respectively,
and are shown in figure \ref{fig_DensityAndSize}(b). 
On the downstream side, the energy drops off quickly, on a length scale that varies very little with $\Rey$.
On the upstream side, the energy drops off more slowly, on a length scale that increases linearly with $\Rey$.
The origin of this scaling is the viscous decay of the streaks on a time scale proportional to $\Rey$, which then is 
translated into a spatial scale proportional to $\Rey$ by the essentially constant advection velocity. In the
case of plane Couette flow, \citet{Brand2014} have been able to determine the slopes from a 
linear stability analysis that confirms this scaling. The case of plane Poiseuille flow is more complicated
because the state is not stationary, and an analytical calculation of the decay rates has not been
possible, yet.

\begin{figure*}
\centering
\includegraphics[]{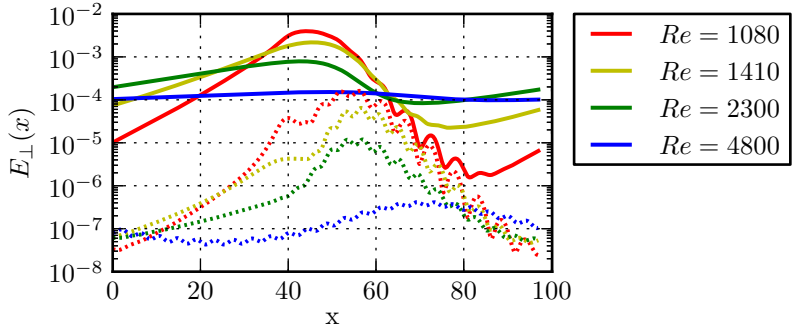}
\caption{Profiles of $PO_{E}$ for increasing Reynolds numbers. The solid lines show the total and the dashed lines the 
cross-flow energy densities. Consistent with the increase in $l_d$, the states become more and more delocalised
with increasing Reynolds number, until they merge into a spatially extended state near Re$=5385$.
The streamwise length of the computational domain is $32\pi$.\label{fig_EdensityRe}}       
\end{figure*}

Although the structure becomes longer with increasing $Re$, the total energy
and also the maximum of the energy density decrease with increasing $\Rey$ (see figure \ref{fig_BifDiag}).
In finite domains the increasing length of the structures will then cause interference 
between head and tail and a loss of localisation.
For instance, continuation of $PO_{E}$ to high $\Rey$ in a box of length $32\pi$ shows that the orbit connects to the 
streamwise extended travelling wave $TW_{E}$ at $Re\approx5385$ with a wavelength of $2.66 \pi$. 
This is documented in figure \ref{fig_EdensityRe} in a plot of the energy densities vs. $\Rey$.
For low $\Rey$ there is a pronounced maximum in the densities, but for increasing $\Rey$ the differences 
decrease and finally at  $Re\approx5380$ the uniform energy density corresponding to the travelling wave 
is obtained. 
Turned the other way round, the localised state arises out of a streamwise long-wavelength 
instability of a travelling wave, very much like the long wavelength instabilities discussed in plane Couette flow \citep{Melnikov2014} or in pipe flow \citep{Chantry2013}. 
 
\section{A streamwise and spanwise localised periodic orbit}
\begin{figure*}
\includegraphics[]{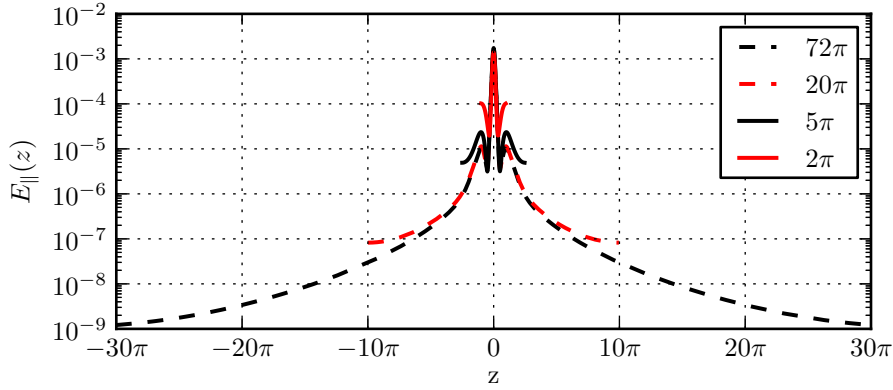}
\caption{Spanwise profiles of the
total energy of the localised states $PO_{E}$ for various spanwise widths $L_{z}$.
\label{fig_EdensityLz}}
\end{figure*}

The periodic orbits found in the domain with $L_{z}=2\pi$ are localised in the streamwise direction.
They show early signs of localisation in the spanwise direction in that the energy density close to the
strong low-speed streak  is much higher than in the region of the high-speed streak.
To obtain periodic orbits that are also localised in the spanwise direction
we continue the periodic orbit in box width. 
For the continuation in $L_{z}$ we fix $Re=2180$ and a length of $64\pi$. As a measure of the state we 
consider the energy density obtained by averaging over the streamwise and normal directions,
\begin{equation}
 E_{\parallel}(z)=\frac{1}{4 L_{x}}\int_{0}^{L_{x}}\int_{-1}^{1} \textbf{u}^{2} \ dx dy .
\end{equation}
This partially averaged energy density depends on the spanwise coordinate only, and is shown in 
figure \ref{fig_EdensityLz}(a) for various widths $L_{z}$. The maximum at $z=0$ corresponds to the position of 
the low-speed streak.  For $L_{z}=2\pi$ the second smaller maximum is the position of the weak high-speed 
streak. Slightly above $L_{z}=2\pi$ the lower maximum splits into two. 
For $L_{z}>5\pi$ the energy density has a very low value over most of the domain, indicating a 
spanwise localised flow structure. 
The largest value of $L_{z}$ which we studied is $72\pi$.  For this domain we use a resolution of 
$N_{x}\times N_{y}\times N_{z}=384 \times 49 \times 1728$.
The doubly-localised solution in this domain keeps the $s_{y}$ symmetry of the state that is 
localised in the streamwise direction only.
The period $T$ of the orbit is $53.578$. After this time the state returns in shape up to a downstream shift and the symmetry operation $s_{z}$.
The group speed of the orbit is $v_{g}=0.8803$.

\begin{figure*}
\centering
\includegraphics[]{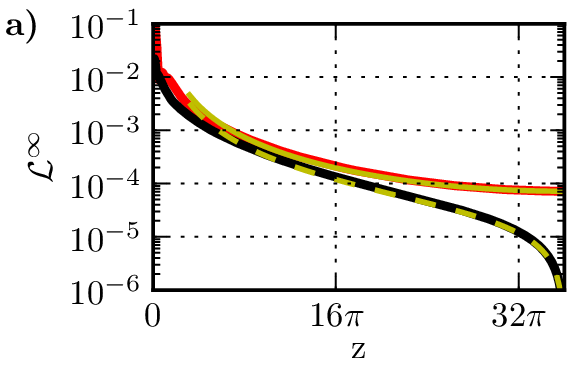}\includegraphics[]{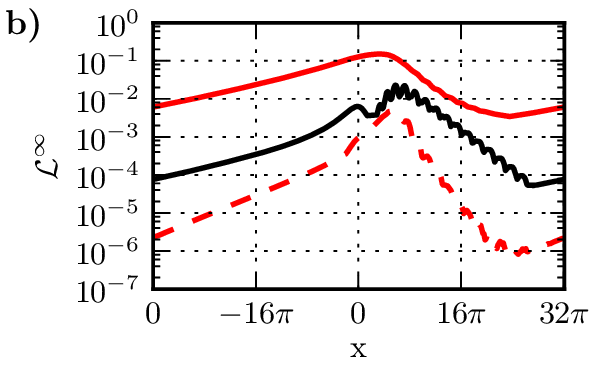}
\caption{In (a) the $\mathcal{L}^{\infty}$-norm  (maximum over y and x) of the streamwise u (solid red), the spanwise w (solid black) velocity component versus spanwise coordinate $z$ are show. 
The yellow lines show fits $A(z^{-2} + (L_{z}-z)^{-2})$ (solid) and $A(z^{-2} - (L_{z}-z)^{-2})$ (dashed).
(b) shows the  $\mathcal{L}^{\infty}$-norms  (maximum over y and z) of the streamwise (solid red), spanwise (black) and wall-normal (dashed red) velocity componet versus streamwise coordinate $x$.}
\label{fig_Linf}       
\end{figure*}

The logarithmic scale in figure \ref{fig_EdensityLz}(a) shows that $E_{\parallel}(z)$ does not drop off 
exponentially in the spanwise direction.
Since the integrated density increases with the length of the turbulent region, a much better
measure is the maximum in velocities along $x$ and $y$ for a fixed spanwise position, 
i.e the $\infty$-norm $\mathcal{L}^{\infty}(u)=\mbox{max}_{x,y}|u(x,y,z)|$, here
given for the streamwise component $u$  \citep{Brand2014}. 
Its values for the streamwise and spanwise component are shown in figure \ref{fig_Linf}(a).
The decay of $\mathcal{L}^{\infty}(u)$ is slower than exponential while $\mathcal{L}^{\infty}(w)$ 
drops off faster. The second part is hidden in  $E_{\parallel}(z)$ because it is swamped by the 
higher values of the streamwise component. The behaviour near $z=L_z$ is clearly influenced by the
boundary conditions: the streamwise component is symmetric under reflection at the boundary,
whereas the spanwise component is antisymmetric and vanishes at the boundary. Taking this into 
account, the figure also shows fits to an algebraic decay with the correct symmetries:
the agreement between the fit and the numerical data indicates that the velocity fields fall off
like $1/z^2$ over the width of the domain.

In the streamwise direction, as documented in figure \ref{fig_Linf}(b), the decay is exponential for the domain sizes studied here.
This agrees with the observations on the partially localised states in section 3, including the asymmetry in the decays in the upstream and downstream direction.

\begin{figure*}
\centering
\includegraphics[]{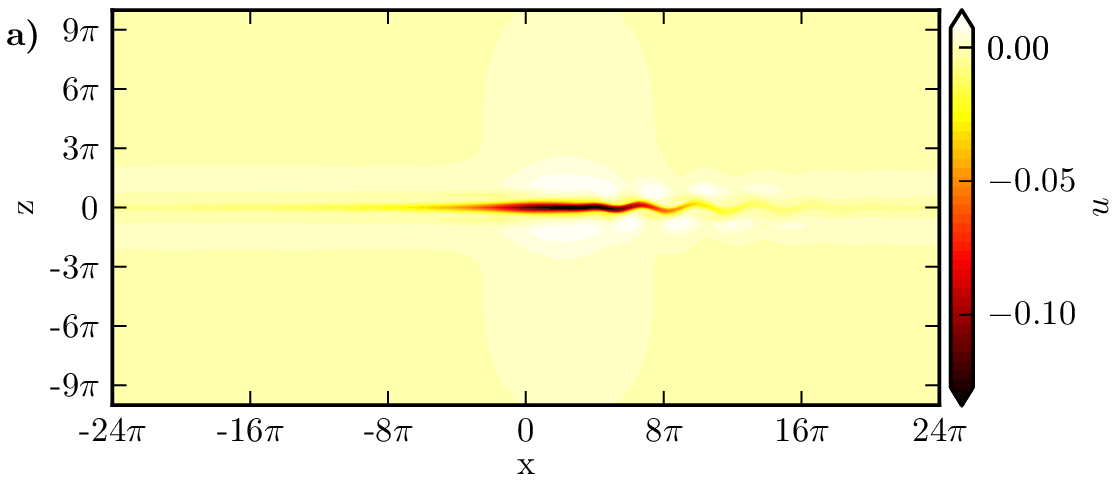}
\includegraphics[]{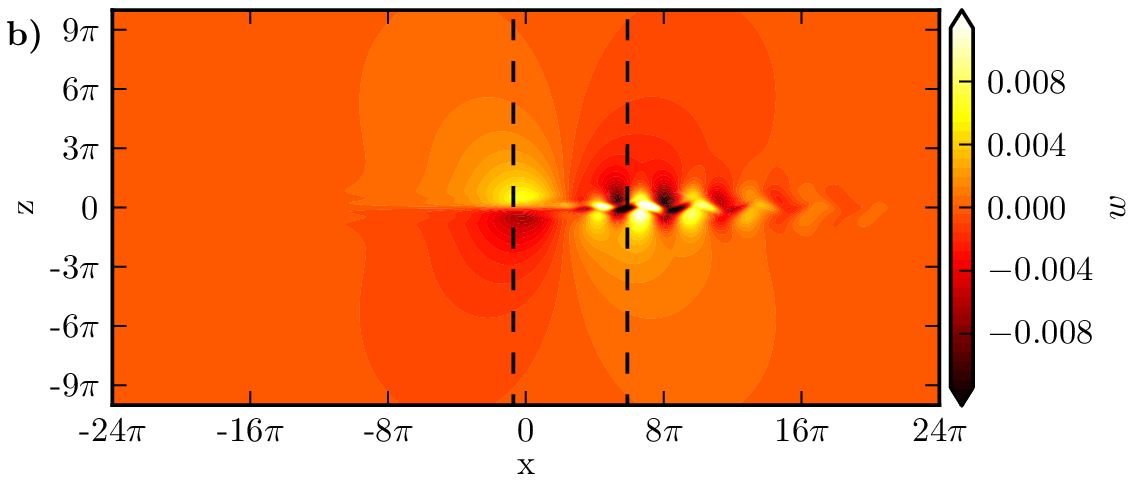}
\caption{Instantaneous streamwise (a) and spanwise (b) velocity in the mid-plane for the doubly-localised periodic orbit at $Re=2180$ in a domain with $L_{x}=64\pi$ and
$L_{z}=72\pi$. The snapshot is for the time of minimal energy during one period. Only the part of the domain that contains the localised flow structure is shown.
The dashed lines in (b) mark the downstream positions of the maxima in the energy density $E_{\perp,c}(x)$.}
\label{fig_PO_XZplaneFullyLoc}       
\end{figure*}

Images of the streamwise and the spanwise velocity fields in the mid-plane 
are shown in figure \ref{fig_PO_XZplaneFullyLoc}. 
The visualisation of the spanwise velocity reveals a large-scale, quadrupolar-like flow field, where the centres
of the left and the right pairs of lobes coincide with intensity maxima of $E_{\perp,c}(x)$.
The quadrupolar shape of the spanwise velocity also exists away from the mid-plane, 
but becomes less distinct close to the walls. 
Given the observation of similar large scale quadrupole flows in turbulent spots in
plane Couette \citep[e.g.][]{Schumacher2001,Lagha2007,Duguet2013,Gibson2014} and plane Poiseuille flow  \citep{Lemoult2013,Lemoult2013b},
one can anticipate that they appear for all structures that are localised in all directions.

We verified that we can trace the doubly-localised solution in the domain with $L_{x}=64\pi$ and $L_{z}=72\pi$ also to lower and higher values of $\Rey$, but
because applying the Newton method to  this large domain is computationally very expensive, we did not perform a complete continuation in Reynolds number.

A stability analysis of the localised state as a function of $L_{z}$ at $Re=2180$ shows that is has two unstable eigenvalues for $L_{z}\geq 6\pi$. 
Therefore,  it is not an attracting state at the laminar-turbulent boundary. Edge tracking calculations starting from the disturbed localised periodic orbit
do not result in a  simple attractor.  Instead, the time evolution of the state is chaotic, but it remains localised \citep{Zammerta}. This behaviour is similar to
what has been seen in large  plane Couette domains \citep{Marinc2010,Schneider2010,Duguet2009}, long pipes \citep{Mellibovsky2009},
or wide domains in the asymptotic suction boundary layer \citep{Khapko2013a}.

\section{Conclusions and Outlook}
We were able to identify a doubly-localised periodic orbit in plane Poiseuille flow.
The orbit was shown to bifurcate from a streamwise extended travelling wave. Together with the other current examples of 
long-wavelength instabilities \citep{Melnikov2014,Chantry2013} we anticipate that many more localised states can be found in bifurcations of the
of spatially extended states that have been identified already \citep{Schmiegel1999,Gibson2009}. Homotopies between plane Poiseuille flow and other flows,
including plane Couette or the asymptotic suction boundary layer, can then reveal connections between these states \citep{Waleffe2003,Kreilos2013}.
More generally, the presence of localised states opens up the path to spatial delocalisation and the development of spatio-temporal
patterns \citep[see e.g.][]{Barkley2005,Avila2011,Tuckerman2014}.
\\

{\bf Acknowledgements} 
We thank John Gibson for providing {\it channelflow} and stimulating exchanges on localisation properties.
We also thank Yohann Duguet and Tobias Kreilos for discussions. 
This work was supported by the Deutsche Forschungsgemeinschaft within FOR 1182.

\bibliographystyle{jfm}


\begin{thebibliography}{0}
\expandafter\ifx\csname natexlab\endcsname\relax\def\natexlab#1{#1}\fi

\end{thebibliography}


\begin{thebibliography}{46}
\expandafter\ifx\csname natexlab\endcsname\relax\def\natexlab#1{#1}\fi

\bibitem[Avila {\em et~al.\/}(2011)Avila, Moxey, de~Lozar, Avila, Barkley \&
  Hof]{Avila2011}
{\sc Avila, K., Moxey, D., de~Lozar, A., Avila, M., Barkley,
  D. \& Hof, B.} 2011 {The onset of turbulence in pipe flow.} {\em
  Science\/} {\bf 333}~(6039), 192--196.

\bibitem[Avila {\em et~al.\/}(2013)Avila, Mellibovsky, Roland \&
  Hof]{Avila2013}
{\sc Avila, M., Mellibovsky, F., Roland, N. \& Hof, B.} 2013
  {Streamwise-localised solutions at the onset of turbulence in pipe flow}.
  {\em Phys. Rev. Lett.\/} {\bf 110}, 224502.

\bibitem[Barkley \& Tuckerman(2005)]{Barkley2005}
{\sc Barkley, D. \& Tuckerman, L.} 2005 {Computational study of turbulent
  laminar patterns in Couette flow}. {\em Phys. Rev. Lett.\/} {\bf 94},
  014502.
  
 \bibitem[Brand \& Gibson(2014)]{Brand2014}
{\sc Brand, E. \& Gibson, J.F.} 2014 {A doubly-localised equilibrium solution of plane Couette flow}. {\em J. Fluid Mech.\/}  {\bf 750}, R1.

\bibitem[Carlson {\em et~al.\/}(1982)Carlson, Widnall \& Peeters]{Carlson1982}
{\sc Carlson, D.~R., Widnall, S.~E. \& Peeters, M.~F.} 1982 {A
  flow-visualization study of transition in plane Poiseuille flow}. {\em J.
  Fluid Mech.\/} {\bf 121}, 487--505.

\bibitem[Chantry {\em et~al.\/}(2014)Chantry, Willis \& Kerswell]{Chantry2013}
{\sc Chantry, M., Willis, A.~P. \& Kerswell, R.~R.} 2013 {The genesis of
  streamwise-localised solutions from globally periodic travelling waves in
  pipe flow}. {\em Phys. Rev. Lett.}  {\bf 112},
  164501.

\bibitem[Dauchot \& Daviaud(1995)]{Dauchot1995}
{\sc Dauchot, O. \& Daviaud, F.} 1995 {Finite amplitude perturbation and spots
  growth mechanism in plane Couette flow}. {\em Phys. Fluids\/} {\bf 7},
  335.

\bibitem[Dijkstra {\em et~al.\/}(2014)Dijkstra, Wubs, Cliffe, Doedel,
  Dragomirescu, Eckhardt, Gelfgat, Hazel, Lucarini, Salinger, Phipps,
  Sanchez-Umbria, Schuttelaars, Tuckerman \& Thiele]{Dijkstra2013}
{\sc Dijkstra, H., {\em et~al.\/}
 } 2014 {Numerical bifurcation methods and their applicaton to fluid
  dynamics: Analysis beyond simulation}. {\em Commun. Comput. Phys.\/} {\bf
  15}, 1--45.

\bibitem[Duguet \& Schlatter(2013)]{Duguet2013}
{\sc Duguet, Y. \& Schlatter, P.} 2013 {Oblique laminar-turbulent interfaces in
  plane shear flows}. {\em Phys. Rev. Lett.\/} {\bf 110}, 034502.

\bibitem[Duguet {\em et~al.\/}(2009)Duguet, Schlatter \&
  Henningson]{Duguet2009}
{\sc Duguet, Y., Schlatter, P. \& Henningson, D.~S.} 2009 {Localised edge
  states in plane Couette flow}. {\em Phys. Fluids\/} {\bf 21}, 111701.

\bibitem[Duguet {\em et~al.\/}(2010)Duguet, Schlatter \&
  Henningson]{Duguet2010b}
{\sc Duguet, Y., Schlatter, P. \& Henningson, D.~S.} 2010 {Formation of
  turbulent patterns near the onset of transition in plane Couette flow}. {\em
  J. Fluid Mech.\/} {\bf 228}, 119-129.

\bibitem[Ehrenstein {\em et~al.\/}(1991)Ehrenstein \& Koch]{Ehrenstein1991}
{\sc Ehrenstein, U. \& Koch, W.} 1991 {Three-dimensional wavelike equilibrium states in plane Poiseuille flow}. {\em J.
  Fluid Mech.\/} {\bf 121}, 111--148.

\bibitem[Gibson(2012)]{J.F.Gibson2012}
{\sc Gibson, J.~F.} 2012 {Channelflow: A spectral Navier-Stokes simulator in
  C++}. {\em Tech. Rep.\/}. U. New Hampshire.

\bibitem[Gibson \& Brand(2014)]{Gibson2014}
{\sc Gibson, J.~F. \& Brand, E.} 2014 {Spanwise-localised solutions of planar
  shear flows}. {\em J. Fluid Mech.\/} {\bf 745}, 25--61.

\bibitem[Gibson {\em et~al.\/}(2009)Gibson, Halcrow \&
  Cvitanovi\'{c}]{Gibson2009}
{\sc Gibson, J.~F., Halcrow, J. \& Cvitanovi\'{c}, P.} 2009 {Equilibrium and
  travelling-wave solutions of plane Couette flow}. {\em J. Fluid Mech.\/} {\bf
  638}, 243--266.

\bibitem[Guennebaud {\em et~al.\/}(2010)Guennebaud, Jacon \& Others]{eigenweb}
{\sc Guennebaud, G., Jacon, B. \& Others} 2010 {Eigen v3}.

\bibitem[Hegseth(1996)]{Hegseth1996}
{\sc Hegseth, J.} 1996 {Turbulent spots in plane Couette flow.} {\em Phys. Rev.
  E\/} {\bf 54}, 4915--4923.

\bibitem[Henningson {\em et~al.\/}(1987)Henningson, Spalart \&
  Kim]{Henningson1987}
{\sc Henningson, D., Spalart, P. \& Kim, J.} 1987 {Numerical simulations of
  turbulent spots in plane Poiseuille and boundary-layer flow}. {\em Phys.
  Fluids\/} {\bf 30}, 2914.

\bibitem[Jeong \& Hussain(1995)]{Hussain95}
{\sc Jeong, J \& Hussain, F} 1995 {On the identification of a vortex}. {\em J.
  Fluid Mech.\/} {\bf 285}, 69--94.
  
\bibitem[Khapko {\em et~al.\/}(2013)Khapko, Kreilos,
  Schlatter, Duguet, Eckhardt \& Henningson]{Khapko2013}
{\sc Khapko, T., Kreilos, T., Schlatter, P., Duguet, Y., Eckhardt, B. \&
  Henningson, D.~S.} 2013 {Localised edge states in the
  asymptotic suction boundary layer}. {\em J. Fluid Mech.\/} {\bf 717}, R6.

\bibitem[Khapko {\em et~al.\/}(2014)Khapko, Duguet,
  Kreilos, Schlatter, Eckhardt \& Henningson]{Khapko2013a}
{\sc Khapko, T., Duguet, Y., Kreilos, T., Schlatter, P., Eckhardt, B. \&
  Henningson, D.~S.} 2014 {Complexity of localised
  coherent structures in a boundary-layer flow}. {\em Eur. Phys. J. E \/} {\bf 37}, 32.


\bibitem[Kreilos \& Eckhardt(2012)]{Kreilos2012}
{\sc Kreilos, T. \& Eckhardt, B.} 2012 {Periodic orbits near onset of chaos in
  plane Couette flow}. {\em Chaos\/} {\bf 22}, 047505.

\bibitem[Kreilos {\em et~al.\/}(2013)Kreilos, Veble, Schneider \&
  Eckhardt]{Kreilos2013}
{\sc Kreilos, T., Veble, G., Schneider, T.~M. \& Eckhardt, B.} 2013 {Edge
  states for the turbulence transition in the asymptotic suction boundary
  layer}. {\em J. Fluid Mech.\/} {\bf 726}, 100--122.
  
\bibitem[Kuznetsov(1998)]{Kuznetsov1998}
{\sc Kuznetsov, Y.A.} 1998 {\em {Elements of applied bifurcation theory}\/}.
  Springer Berlin / Heidelberg.

\bibitem[Lagha \& Manneville(2007)]{Lagha2007}
{\sc Lagha, M. \& Manneville, P.} 2007 {Modeling of plane Couette flow. I.
  Large scale flow around turbulent spots}. {\em Phys. Fluids\/} {\bf 19},
  094105.

\bibitem[Lemoult {\em et~al.\/}(2013)Lemoult, Aider \&
  Wesfreid]{Lemoult2013}
{\sc Lemoult, G., Aider, J.-L. \& Wesfreid, J.~E.} 2013{\natexlab{{\em a\/}}}
  {Turbulent spots in a channel: large-scale flow and self-sustainability}.
  {\em J. Fluid Mech.\/} {\bf 731}, R1.

\bibitem[Lemoult {\em et~al.\/}(2014)Lemoult, Gumowski,
  Aider \& Wesfreid]{Lemoult2013b}
{\sc Lemoult, G., Gumowski, K., Aider, J.-L. \& Wesfreid, J.~E.}
  2014 {Turbulent spots in channel : an experimental
  study Large-scale flow, inner structure and low order model}. {\em Eur. Phys. J. E \/} {\bf 37}, 25.

\bibitem[Lundbladh \& Johansson(1991)]{Lundbladh1991}
{\sc Lundbladh, A. \& Johansson, A. V.} 1991 {Direct simulation of turbulent spots
  in plane Couette flow}. {\em J. Fluid Mech.\/} {\bf 229}.

\bibitem[Manneville(2009)Manneville]{Manneville2009}
{\sc Manneville, P.} 2009
  {Spatiotemporal perspective on the decay of turbulence in wall-bounded flows}. {\em Phys. Rev. E\/} {\bf
  79}, 025301.

\bibitem[Marinc {\em et~al.\/}(2010)Marinc, Schneider \& Eckhardt]{Marinc2010}
{\sc Marinc, D., Schneider, T.~M. \& Eckhardt, B.} 2010 {localised edge states
  for the transition to turbulence in shear ﬂows}. In {\em Seventh IUTAM
  Symp. Laminar-Turbulent Transit.\/} (ed. Philipp Schlatter \& Dan~S.
  Henningson), {\em IUTAM Bookseries\/}, vol.~18, pp. 253--258. Dordrecht:
  Springer Netherlands.

\bibitem[Mellibovsky {\em et~al.\/}(2009)Mellibovsky, Meseguer, Schneider \&
  Eckhardt]{Mellibovsky2009}
{\sc Mellibovsky, F., Meseguer, A., Schneider, T. \& Eckhardt, B.} 2009
  {Transition in localised Pipe Flow Turbulence}. {\em Phys. Rev. Lett.\/} {\bf
  103}, 054502.

\bibitem[Melnikov {\em et~al.\/}(2014)Melnikov, Kreilos \&
  Eckhardt]{Melnikov2014}
{\sc Melnikov, K., Kreilos, T. \& Eckhardt, B.} 2014 {Long wavelength
  instability of coherent structures in plane Couette flow}. {\em Phys. Rev.
  E\/} {\bf 89}, 043088.

\bibitem[Moxey \& Barkley(2010)]{Moxey2010}
{\sc Moxey, D. \& Barkley, D.} 2010 {Distinct large-scale
  turbulent-laminar states in transitional pipe flow.} {\em Proc. Natl. Acad.
  Sci. U. S. A.\/} {\bf 107}, 8091--8096.

\bibitem[Nagata(1990)Nagata]{Nagata1990}
{\sc Nagata, M.} 1990 {Three-dimensional finite-amplitude solutions in plane Couette flow: bifurcation from infinity}. {\em J.
  Fluid Mech.\/} {\bf 217}, 519--527.
  
  \bibitem[Nagata(1997)Nagata]{Nagata1997}
{\sc Nagata, M.} 1997 {Three-dimensional traveling-wave solutions in plane Couette flow}. {\em Phys. Rev.
  E\/} {\bf 55}, 2023--2025.

\bibitem[Nagata \& Deguchi(2013)]{Nagata2013a}
{\sc Nagata, M. \& Deguchi, K.} 2013 {Mirror-symmetric exact coherent states in
  plane Poiseuille flow}. {\em J. Fluid Mech\/} {\bf 735}, R4.

\bibitem[Orszag(1971)]{Orszag1971}
{\sc Orszag, S.~A.} 1971 {Accurate solution of the Orr--Sommerfeld stability
  equation}. {\em J. Fluid Mech.\/} {\bf 50}, 689--703.

\bibitem[Price {\em et~al.\/}(1993)Price, Brachet \& Pomeau]{Price1993}
{\sc Price, T., Brachet, M. \& Pomeau, Y.} 1993 {Numerical characterization of
  localised solutions in plane Poiseuille flow}. {\em Phys. Fluids A Fluid
  Dyn.\/} {\bf 5}, 762.
  
\bibitem[Schmiegel(1999)]{Schmiegel1999}
{\sc Schmiegel, Armin} 1999 {Transition to turbulence in linearly stable shear
  flows}. Phd thesis, Marburg.


\bibitem[Schneider {\em et~al.\/}(2008)Schneider, Gibson, Lagha, {De Lillo} \&
  Eckhardt]{Schneider2008}
{\sc Schneider, T.~M., Gibson, J.~F., Lagha, M., {De Lillo}, F. \& Eckhardt,
  B.} 2008 {Laminar-turbulent boundary in plane Couette flow}. {\em Phys. Rev.
  E\/} {\bf 78}, 037301.
  
\bibitem[Schneider {\em et~al.\/}(2010{\natexlab{{\em a\/}}})Schneider, Gibson
  \& Burke]{Schneider2010a}
{\sc Schneider, T.~M., Gibson, J.~F. \& Burke, J.} 2010{\natexlab{{\em a\/}}}
  {Snakes and ladders: Localised solutions of plane Couette flow}. {\em Phys.
  Rev. Lett.\/} {\bf 104}, 104501.

\bibitem[Schneider {\em et~al.\/}(2010{\natexlab{{\em b\/}}})Schneider, Marinc
  \& Eckhardt]{Schneider2010}
{\sc Schneider, T.~M., Marinc, D. \& Eckhardt, B.} 2010{\natexlab{{\em b\/}}}
  {localised edge states nucleate turbulence in extended plane Couette cells}.
  {\em J. Fluid Mech.\/} {\bf 646}, 441.

\bibitem[Schumacher \& Eckhardt(2001)]{Schumacher2001}
{\sc Schumacher, J. \& {Eckhardt}, B.} 2001 {Evolution of turbulent spots in a
  parallel shear flow}. {\em Phys. Rev. E\/} {\bf 63}, 046307.

\bibitem[Skufca {\em et~al.\/}(2006)Skufca, Yorke \& Eckhardt]{Skufca2006}
{\sc Skufca, J., Yorke, J. \& Eckhardt, B.} 2006 {Edge of chaos in a parallel
  shear flow}. {\em Phys. Rev. Lett.\/} {\bf 96}, 174101.
  

\bibitem[Toh \& Itano(2003)]{Toh2003}
{\sc Toh, S. \& Itano, T.} 2003 {A periodic-like solution in channel flow}.
  {\em J. Fluid Mech.\/} {\bf 481}, 67--76.
  
 \bibitem[Tuckerman {\em et~al.\/}(2014)Tuckerman, Kreilos, Schrobsdorff, Schneider \&Gibson]{Tuckerman2014}
{\sc Tuckerman, L.,Kreilos, T., Schrobsdorff, H., Schneider, T.M.  \& Gibson, J. F.} 2014 {Turbulent-laminar patterns in plane Poiseuille flow}.
  {\em  arXiv1312.6783\/}

\bibitem[Viswanath(2007)]{Viswanath2007}
{\sc Viswanath, D.} 2007 {Recurrent motions within plane Couette turbulence}.
  {\em J. Fluid Mech.\/} {\bf 580}, 339.

  
\bibitem[Waleffe(2001)]{Waleffe2001}
{\sc Waleffe, F.} 2001 {Exact coherent structures in channel flow}. {\em J. Fluid Mech.\/} {\bf 435}, 93--102.

\bibitem[Waleffe(2003)]{Waleffe2003}
{\sc Waleffe, F.} 2003 {Homotopy of exact coherent structures in plane shear
  flows}. {\em Phys. Fluids\/} {\bf 15}, 1517.

\bibitem[Wang {\em et~al.\/}(2007)Wang, Gibson \& Waleffe]{Wang2007}
{\sc Wang, J., Gibson, J.~F. \& Waleffe, F.} 2007 {Lower branch coherent
  states in shear flows: Transition and control}. {\em Phys. Rev. Lett.\/} {\bf
  98}, 204501.

\bibitem[Zammert \& Eckhardt(2014)]{Zammert2013}
{\sc Zammert, S. \& Eckhardt, B.} 2014
  {Periodically bursting edge states in plane Poiseuille flow}. {\em Fluid. Dyn. Res.\/} {\bf 46}, 041419.
  
\bibitem[Zammert \& Eckhardt(2014{\natexlab{{\em b\/}}})]{Zammerta}
{\sc Zammert, S. \& Eckhardt, B.} 2014{\natexlab{{\em b\/}}} {A spotlike edge state in plane
  Poiseuille flow}. {\em Proc. Appl. Math. Mech.\/}
  submitted.

\end{thebibliography}

\end{document}